# Superconductivity in Ternary Zirconium Telluride $Zr_6RuTe_2$

Kosuke Yuchi*, Haruka Matsumoto, Daisuke Nishio-Hamane, Kodai Moriyama, Keita Kojima, Ryutaro Okuma, Jun-ichi Yamaura, and Yoshihiko Okamoto

*Institute for Solid State Physics, University of Tokyo, Kashiwa 277-8581, Japan*

$Zr_6CoAl_2$-type $Zr_6RuTe_2$ is found to show bulk superconductivity below the superconducting transition temperature $T_c$ = 1.1 K, according to the electrical resistivity, magnetization, and heat capacity measurements using synthesized polycrystalline samples. This $T_c$ exceeds that of $Zr_6MTe_2$ compounds in which M is other transition metals, indicating that M = Ru is favorable for superconductivity in $Zr_6CoAl_2$-type $Zr_6MX_2$.

Recently, several new superconductors have been discovered in $Zr_6CoAl_2$-type $A_6MX_2$ compounds exhibiting a noncentrosymmetric $P$–$62m$ space group. This finding was initiated by $Sc_6MTe_2$, which was reported to exhibit bulk superconductivity for seven transition-metal elements M,[1] followed by several $Zr_6MX_2$ compounds.[2–4] A unique feature of this superconductor family is M dependence of the superconducting transition temperature, $T_c$. In $Sc_6MTe_2$ and $Zr_6MTe_2$ compounds, the highest $T_c$ is recorded for M = Fe ($T_c$ = 4.7 K for $Sc_6FeTe_2$ and 0.76 K for $Zr_6FeTe_2$). In contrast, $Zr_6MBi_2$ superconductors showed a different M dependence of $T_c$. $Zr_6RuBi_2$ exhibits a significantly elevated $T_c$ of 4.9 K compared to $Zr_6FeBi_2$ ($T_c$ = 1.4 K). Although the Fe 3$d$ electrons are suggested to contribute to the realization of a high $T_c$ in $Sc_6FeTe_2$,[1,5] the highest $T_c$ of $Zr_6RuBi_2$ among all $A_6MX_2$ compounds indicates a complex chemical trend in these superconductors, highlighting the importance of continued exploration of new $A_6MX_2$ superconductors. Recently, superconducting properties other than $T_c$ have also been revealed using microscopic probes such as NMR and μSR.[6,7]

In this short note, we focus on $Zr_6RuTe_2$, which crystallizes in the hexagonal $Zr_6CoAl_2$-type structure.[8] There have been no previous reports on its electronic properties. We prepared polycrystalline samples of $Zr_6RuTe_2$ and found that it is a bulk superconductor with $T_c$ = 1.1 K.

Polycrystalline samples of $Zr_6RuTe_2$ were synthesized and characterized using the same procedure as described in Ref. 4. A 6:1:2.1 molar ratio of Zr chips (99.9%, RARE METALLIC), Ru powder (99.95%, RARE METALLIC), and Te powder (99.99%, RARE METALLIC) was arc-melted, and the resulting button was annealed at 1373 K for 216 h. Powder X-ray diffraction (XRD) measurements of $Zr_6RuTe_2$ polycrystalline samples were conducted at BL02B2 in SPring-8. The X-ray wavelength was λ = 0.354978 Å. Structural analysis was performed utilizing Rietveld method on JANA2006 software.[9] The XRD pattern at room temperature is accurately represented by the hexagonal $Zr_6CoAl_2$-type structural model, with lattice parameters $a$ = 7.8203(3) Å and $c$ = 3.6220(2) Å, consistent with the previous study.[8] The reliability factors of the refinement are $R_{wp}$ = 7.92%, $R_p$ = 5.80%, and $S$ = 3.13. Microstructural observations and chemical analyses were carried out using a scanning electron microscope (SEM; JEOL IT-100) equipped with an attachment for energy-dispersive X-ray spectroscopy (EDX; 15 kV, 0.8 nA, 1 μm beam diameter). The inset of Fig. 1(a) shows that the $Zr_6RuTe_2$ polycrystalline sample comprises not only the $Zr_6RuTe_2$ phase (region marked A), representing 89% of the total area, but also the ZrRu and Zr phases (region marked B).[10] The volume fraction of the $Zr_6RuTe_2$ phase was significantly increased by annealing. The chemical composition of the $Zr_6RuTe_2$ phase, as determined by EDX, was $Zr_{6.00(2)}Ru_{0.93(2)}Te_{2.07(1)}$. Considering the small degree of nonstoichiometry, the chemical formula of the $Zr_6RuTe_2$ phase is expressed as $Zr_6RuTe_2$ in this study. Electrical resistivity and heat capacity were measured using a Physical Property Measurement System (Quantum Design). Magnetization measurements were performed using an MPMS-3 (Quantum Design).

Figures 1(a) and 1(b) display the temperature dependence of the electrical resistivity, ρ, of the $Zr_6RuTe_2$ polycrystalline sample. As shown in Fig. 1(a), $Zr_6RuTe_2$ exhibits metallic behavior. The residual resistivity ratio RRR = $ρ_{300\,K}/ρ_0$ was estimated to be 6.9, where $ρ_{300\,K}$ and $ρ_0$ denote the ρ at 300 K and the residual resistivity, respectively. As the temperature was further lowered, ρ slightly decreased at $T^*$ = 1.8 K and became nearly constant at approximately 1.5 K, as shown in Fig. 1(b). Subsequently, ρ dropped sharply to zero between 1.3 and 1.0 K. This sharp drop was suppressed to lower temperatures under applied magnetic fields, strongly suggesting a superconducting transition. As shown in Fig. 1(c), the zero-field-cooled (ZFC) and field-cooled (FC) magnetization data of the $Zr_6RuTe_2$ polycrystalline sample under a magnetic field of 10 Oe exhibited a pronounced diamagnetic signal below 1.1 K with a subdued diamagnetic signal below 1.8 K. The shielding fraction at 0.4 K was eval-uated to be 165%, assuming the single phase of $Zr_6RuTe_2$. The fraction well beyond 100% is probably due to a demagnetization effect.[11–13] Figure 1(d) shows the heat capacity divided by temperature, $C/T$, which increases with decreasing temperature below 1.2 K and exhibited a peak at 0.8 K. These results clearly indicate that the $Zr_6RuTe_2$ polycrystalline samples exhibit bulk superconductivity. Based on the midpoint of the resistivity drop and the onset of the strong magnetization drop below 1.3 K, $T_c$ was determined to be 1.1 K.

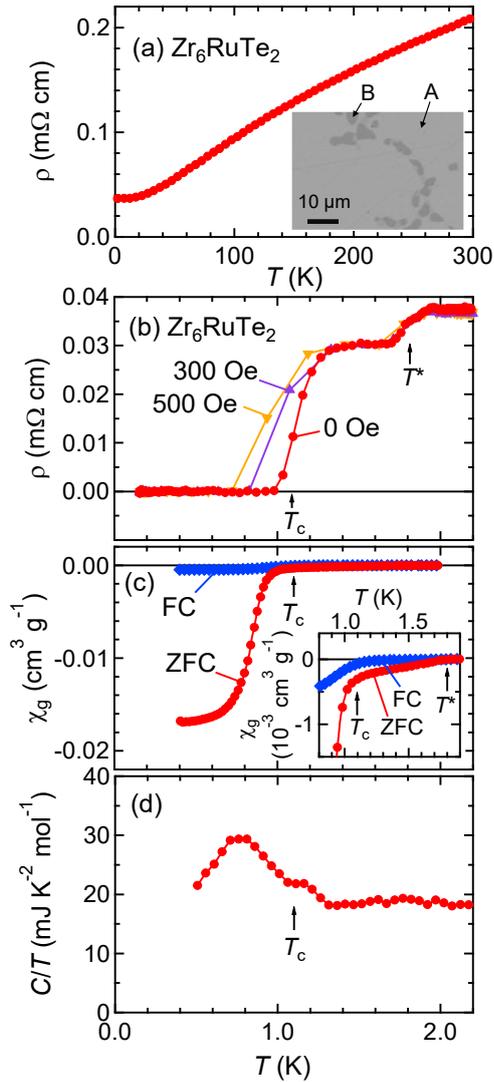

Figure 1. Electronic properties of polycrystalline samples of $Zr_6RuTe_2$. (a) Temperature dependence of electrical resistivity above 2 K. The inset shows SEM image of a $Zr_6RuTe_2$ polycrystalline sample. (b) Temperature dependences of electrical resistivity measured under magnetic fields of 0, 300, and 500 Oe. (c) Temperature dependences of FC and ZFC magnetic susceptibility measured under a magnetic field of 10 Oe. The inset shows an enlarged view of the main panel. (d) Temperature dependence of heat capacity divided by temperature, $C/T$, measured in zero magnetic field. The $C/T$ values were calculated based on the assumption that the sample consists of a single phase of $Zr_6RuTe_2$.

The weak decrease of $\rho$ at $T^*$ is most likely due to the superconductivity of an impurity phase, because a small diamagnetic signal is observed between $T^*$ and $T_c$ in the magnetic susceptibility data and no anomaly appears near $T^*$ in the $C/T$ data. The $Zr_6RuTe_2$ polycrystalline samples used in this study contained approximately 10% ZrRu and Zr impurity phases. However, these impurity phases do not exhibit superconductivity at $T^*$. Zr is a known superconductor with $T_c = 0.6$ K.[14,15] We synthesized a ZrRu polycrystalline sample and found that the ZrRu sample does not exhibit superconductivity above 0.2 K. Therefore, the origin of the weak decrease of $\rho$ at $T^*$ remains as an open question.

The observed $T_c = 1.1$ K of $Zr_6RuTe_2$ is the highest among $Zr_6MTe_2$ compounds, although it is lower than the $T_c$ of other Ru-based compounds, such as $Zr_6RuBi_2$ and $Sc_6RuTe_2$.[1,4] In the $Zr_6MTe_2$ series, compounds with M = Fe and Co exhibit $T_c = 0.76$ K and 0.13 K, respectively, whereas those with M = Cr and Mn do not show zero resistivity above 0.1 K.[2] The M dependence of $T_c$ of $Zr_6MTe_2$ is similar to that of $Zr_6MBi_2$, where $Zr_6RuBi_2$ has a significantly elevated $T_c$ compared to $Zr_6FeBi_2$.[4] This M dependence differs from that of $Sc_6MTe_2$. $Sc_6FeTe_2$ exhibits the highest $T_c$ among the eight $Sc_6MTe_2$ compounds synthesized thus far. In $Sc_6FeTe_2$, both the Sc $3d$ and Fe $3d$ electrons have a large contribution to the density of states at the Fermi energy $E_F$, which most likely plays an important role in the realization of high $T_c$.[1,5] However, in $Sc_6MTe_2$ with M = $4d$ and $5d$ transition metals, Sc $3d$ electrons dominate the electronic state at $E_F$, and they share a lower $T_c$ of approximately 2 K. In contrast, in the Zr series, both $Zr_6RuTe_2$ and $Zr_6RuBi_2$ exhibit the highest $T_c$ in the $Zr_6MTe_2$ and $Zr_6MBi_2$ compounds, respectively. The first principles calculations of $Zr_6RuBi_2$ show that the contribution of the Ru $4d$ electrons at $E_F$ is comparable to that of the Fe $3d$ electrons in $Sc_6FeTe_2$.[16] The large contribution of the Ru $4d$ electrons at $E_F$ in $Zr_6RuX_2$, which is in contrast to $Sc_6RuTe_2$, suggests that the Ru $4d$ electrons may play an important role in achieving the high $T_c$ values. We hope that further experimental and theoretical studies will elucidate the role of Ru $4d$ electrons in the superconductivity of $Zr_6RuX_2$, which will lead to the discovery of even higher $T_c$ and unconventional superconductors in the $A_6MX_2$ family.

**Acknowledgments**

The authors are grateful to Y. Yamakawa, R. Ishii, and Z. Hiroi for helpful discussions. This study was supported by the JSPS KAKENHI (Grant Nos. 23H01831 and 23K26524) and JST ASPIRE (Grant No. JPMJAP2314). The powder XRD experiments were conducted at SPring-8 (Proposal No. 2024B2007) in Hyogo, Japan.

*email: yuchi@issp.u-tokyo.ac.jp